\documentclass{iau}
\usepackage{graphicx}

\title[UVMag] 
{UVMag: a UV and optical spectropolarimeter for stellar physics}

\author[Neiner et al.]   
{Coralie Neiner$^1$, 
Pascal Petit$^{2}$, 
Laurent Par\`es$^{2}$
\and the UVMag consortium
}

\affiliation{$^1$LESIA, UMR 8109 du CNRS, Observatoire de Paris, UPMC, Univ. Paris Diderot, 5 place Jules Janssen, 92195 Meudon Cedex, France\\ email: {\tt coralie.neiner@obspm.fr}\\[\affilskip]
$^2$IRAP, CNRS, UPS-OMP, 14 Avenue Edouard Belin, 31400 Toulouse, France}

\pubyear{2013}
\volume{302}  
\pagerange{119--126}
\jname{Magnetic fields throughout stellar evolution}
\editors{A.C. Editor, B.D. Editor \& C.E. Editor, eds.}
\begin{document}

\maketitle

\begin{abstract}
UVMag is a space project currently under R\&D study. It consists in a
medium-size telescope equipped with a spectropolarimeter to observe in the UV
and optical wavelength domains simultaneously. Its first goal is to obtain time
series of selected magnetic stars over their rotation period, to study them from
their surface to their environment, in particular their wind and magnetospheres.
As the star rotates it will be possible to reconstruct 3D maps of the star and
its surroundings. The second goal of UVMag is to obtain two observations of a
large sample of stars to construct a new database of UV and optical
spectropolarimetric measurements.
\end{abstract}

\firstsection 
\section{Science drivers}

Research on stellar magnetism has been progressing very fast from the ground in
the last decade, but we are missing information on stellar wind and
magnetospheres because there is no UV spectrograph currently available for long
fractions of time. To reconstruct the magnetospheres, we need to obtain
simultaneous UV and optical spectropolarimetry, continuously over several
stellar rotation periods. Of course, the UV domain requires a space mission.
Therefore the UVMag consortium proposes to build a dedicated M-size space
mission with a telescope of 1.3 m and UV and optical spectropolarimetric
capabilities (see http://lesia.obspm.fr/UVMag). 

The UV and visible spectropolarimeter will provide a very powerful and unique
tool to study most aspects of stellar physics in general and in particular for
stellar formation, structure and evolution as well as for stellar environment.
For example we plan to study how fossil magnetic fields confines the wind of
massive stars and influences wind clumping, how magnetic interactions impact
binary stars, how a solar dynamo impacts its planets and how it evolves, how
magnetic field, wind and mass-loss influence the late stages of stellar
evolution, in which conditions a magnetic dynamo develops, how the angular
momentum of stars evolves, how small-scale and large-scale stellar dynamos work
and how their cycles influence their environment, what explains the diversity of
magnetic properties in M dwarfs, what causes the segregation of tepid stars in
two categories: those with sub-Gauss magnetic fields and those with fields above
a few hundreds Gauss, what are the timescales over which magnetospheric
accretion stops in PMS stars, etc. These questions will be answered by observing
all types of stars: massive stars, giants and supergiants, chemically peculiar
stars, pre-main sequence stars, cool stars, solar twins, M dwarfs, AGB and
post-AGB stars, binaries, etc. Additional possible science includes the study of
the ISM, white dwarfs, novae, exoplanets, atomic physics,...

\section{UV and optical spectropolarimeter}

The spectropolarimeter should ideally cover the full wavelength range from 90 to
1000 nm and at least the most important lines in the domains 117-320 nm and
390-870 nm. Polarisation should be measured at least in Stokes V (circular
polarisation) in spectral lines, but the aim is to measure all Stokes QUV
parameters (circular and linear polarisation) in the lines and continuum. A high
spectral resolution is required, at least 25000 in the UV domain and at least
35000 in the optical, with a goal of 80000 to 100000. The signal-to-noise should
be above 100.

Spectroscopy with these specifications in the UV and optical domains is
relatively easy to achieve with today's technology and detectors. However, (1)
high-resolution spectropolarimetry of stars has never been obtained from space;
(2) optical spectropolarimeters available on the ground are large; and (3) it is
very important to keep the instrumental polarisation at a low level. Therefore
we have started a R\&D program to study a space UV+optical spectropolarimeter.
Our study is based on existing ground-based spectropolarimeters, such as
ESPaDOnS or Narval, and new spectropolarimetric techniques proposed in the
literature (e.g. \cite[Sparks et al. 2012]{sparks2012}).

\section{Observing program}

UVMag will observe all types of stars in the magnitude range at least V=3-10. 
The observing program includes two parts: (1) $\sim$50 stars will be observed
over 2 full rotational cycles with high cadence in order to study them in great
details and reconstruct 3D maps of their surface and environment. In addition,
the solar-like stars among those will be re-observed every year to study their
activity cycle; and (2) two spectropolarimetric measurements of $\sim$4000 stars
will be obtained to provide information on their magnetic field, wind and
environment. This will form a statistical survey and provide input for stellar
modelling. The acquisition of the data for these two programs will take 4
years. 

\section{Conclusions}

The UVMag consortium has set the basic requirements for a M-size (1.3 m) space
mission to study the magnetospheres and winds of all types of stars. This is the
next step to progress on the characterisation and modelling of stellar
environments, as well as on important questions regarding stellar formation,
structure and evolution. Simultaneous UV and optical spectropolarimetry over
long periods of time is indeed the only way to comprehend the full interaction
between various physical processes such as the stellar magnetic field and
stellar wind. A R\&D study is ongoing for the instrument. The M-size mission
will be proposed at ESA. A L-size mission (4-8 meter telescope) is also
considered with the UVMag UV and optical spectropolarimeter as part of a series
of instrument, e.g. on EUVO \cite[(G\'omez de Castro et al., 2013)]{euvo2013}.

\begin{acknowledgments}
The UVMag R\&D program is funded by the French space agency CNES.
\end{acknowledgments}

\end{document}